\documentclass{aastex}
\usepackage{emulateapj5}

\newcommand{\iz}{I~Zw~18}
\newcommand{\kms}{km~s$^{-1}$}

\begin{document}

\title{No diffuse H$_2$ in the metal deficient galaxy I~Z\lowercase{w}~18}

\author{A.~Vidal-Madjar\altaffilmark{1}, 
D.~Kunth\altaffilmark{1}, 
A.~Lecavelier des Etangs\altaffilmark{1}, 
J.~Lequeux\altaffilmark{2},
M.~Andr\'e\altaffilmark{3}, 
L.~BenJaffel\altaffilmark{1}, 
R.~Ferlet\altaffilmark{1}, 
G.~H\'ebrard\altaffilmark{1}, 
J.C.~Howk\altaffilmark{3},
J.W.~Kruk\altaffilmark{3},
M.~Lemoine\altaffilmark{4},
H.W.~Moos\altaffilmark{3},
K.C.~Roth\altaffilmark{3},
G.~Sonneborn\altaffilmark{5},
and D.G.~York\altaffilmark{6}}
\altaffiltext{1}{Institut d'Astrophysique de Paris, CNRS, 98 bis bld Arago,
F-75014 Paris, France}
\altaffiltext{2}{DEMIRM, Observatoire de Paris, 61 rue de l'observatoire, 
F-75014 Paris}
\altaffiltext{3}{Department of Physics and Astronomy, Johns Hopkins 
University, Baltimore, MD 21218, USA}
\altaffiltext{4}{DARC, UMR--8629 CNRS, Observatoire de Paris-Meudon, 
F-92195 Meudon, France}
\altaffiltext{5}{Laboratory for Astronomy and Solar Physics, NASA/GSFC, 
Code 681, Greenbelt, MD 20771, USA}
\altaffiltext{6}{Dept. of Astronomy \& Astrophysics, University of Chicago, 
Chicago, IL 60637, USA}

\begin{abstract}

The metal deficient starburst galaxy \iz\ has been observed with 
{\it FUSE} in a search for H$_2$ molecules.
The spectrum obtained with an aperture covering the full galaxy shows 
no absorption lines of diffuse H$_2$ at the radial velocity of the galaxy.
The upper limit for the diffuse H$_2$ column density is found to 
be very low:
$N({\rm H}_2)\la 10^{15}$~cm$^{-2}$ ($10\,\sigma$), unlike our Galaxy 
where H$_2$ is generally present for even low H\,{\sc i} column densities.
Although the H\,{\sc i} column density is here as high as 
N(H\,{\sc i})$\approx 2\times 10^{21}$cm$^{-2}$,
we observe 2N(H$_2$)/N(H\,{\sc i})$\ll 10^{-6}$.
We cannot exclude the possibility that some H$_2$ could be in very dense, 
small and discrete clumps which cannot be detected with the present 
observation. However, the remarkable absence of diffuse H$_2$ 
in this metal-poor galaxy can be explained 
by the low abundance of dust grains (needed to form this molecule from 
H-atoms), the high ultraviolet flux and the low density of the H\,{\sc i} 
cloud surrounding the star--forming regions.
Thus having eliminated diffuse H$_2$ as
a significant contributor to
the total mass, 
it appears that the gas of the galaxy is dominated by H\,{\sc i}, and 
that the high dynamical mass is not composed of cold and 
diffuse baryonic dark matter.

\end{abstract}


\section{Introduction}
\iz\ (Mkn 116) is a dwarf blue compact galaxy presently experiencing a strong
burst of star formation which has produced a pair of bright 
H\,{\sc II} regions.
This galaxy has the smallest known abundance of heavy 
elements as derived from the ionized 
gaseous component. 
Its oxygen abundance is only $\sim$ 1/50 of that of the Sun.
The distribution and kinematics of neutral hydrogen derived from
aperture synthesis observations have been discussed in several works.
These works have derived H\,{\sc i} masses in the range 
$3$---$7\times 10^{7}$M$_\sun$ and
dynamical masses in the range $3$---$9\times 10^{8}$M$_\sun$ 
(Lequeux \& Viallefond 1980; Viallefond et al. 1987; van Zee et al. 1998).
Van Zee et al. (1998)
have emphasized the complexity of the H\,{\sc i} velocity fields, while
Martin (1996) and Petrosian et al. (1997) have discussed the ionized
component.
It has been suggested that objects with localized massive star
formation surrounded by large H\,{\sc i} envelopes might contain a
significant reservoir of molecular hydrogen. Such material could
represent a significant fraction of the dark matter (Lequeux \&
Viallefond 1980). Attempts to detect CO in H\,{\sc II} galaxies have
so far been unsuccessful (Combes 1986; Young et al. 1986; Arnault et
al. 1988; Sage et al. 1992; Isra\"el et al. 1995; Gondhalekar et
al. 1998). This lack of detection does not necessarily imply a lack of
H$_2$: the CO excitation could be lower than for molecular clouds in
our Galaxy or CO might be more photodissociated than H$_2$; but
perhaps most importantly, C and O are highly underabundant in these
metal deficient galaxies. The lack of detectable molecular material
also has other important implications for galaxies like \iz . Given
the chemically unevolved nature of \iz\ and its lack of organized gas
dynamics and/or spiral arms, it is unclear where and how this galaxy
formed the molecular gas thought to be required to form the current
generation of young stars.

Therefore, we observed \iz\ with the Far Ultraviolet Spectroscopic
Explorer ({\it FUSE}, Moos et al.~2000) with the aim of detecting cold
molecular hydrogen lines in absorption against the stellar continuum
of blue massive stellar clusters.
In Sect.~2 we describe the observations and the data analysis, the results
are discussed in Sect.~3

\section{Data analysis}

\iz\ has been observed for 31600 seconds on November 28, 1999 with {\it FUSE}
through the two LiF channels ($\sim$ 980 to
1187 \AA). The large entrance aperture (30\arcsec x 30\arcsec ) 
has been used, fully covering the galaxy.
The data have been processed with the pipeline version 1.5.
The spectral resolution is defined by both the instrument 
and the size of the galaxy (10\arcsec). We find a resolution of 
about $\lambda / \Delta
\lambda \sim 10000 $ with a S/N ratio of $\sim 10$ per resolution element.

Many absorption lines are clearly detected. They correspond to 
three main components at different radial velocities: 
$-260$~\kms, $-100$~\kms, and $650$~\kms. These can easily be identified 
respectively with
the known high velocity cloud at $-160$~\kms, the clouds within 
the Galaxy expected at low radial velocity, and \iz\ itself with a redshift
of $750$~\kms. 
We thus conclude that there is a systematic wavelength shift in the 
whole spectrum corresponding to a blueshift of about 100~km~s$^{-1}$.
This systematic wavelength shift 
can be explained by the preliminary wavelength calibration of {\it
FUSE} and by the position of the target possibly off the center of the slit.
All the velocities quoted below refer to the observed velocities
corrected by this systematic effect assumed to be exactly 
100~km~s$^{-1}$.

\begin{figure*}
\plotone{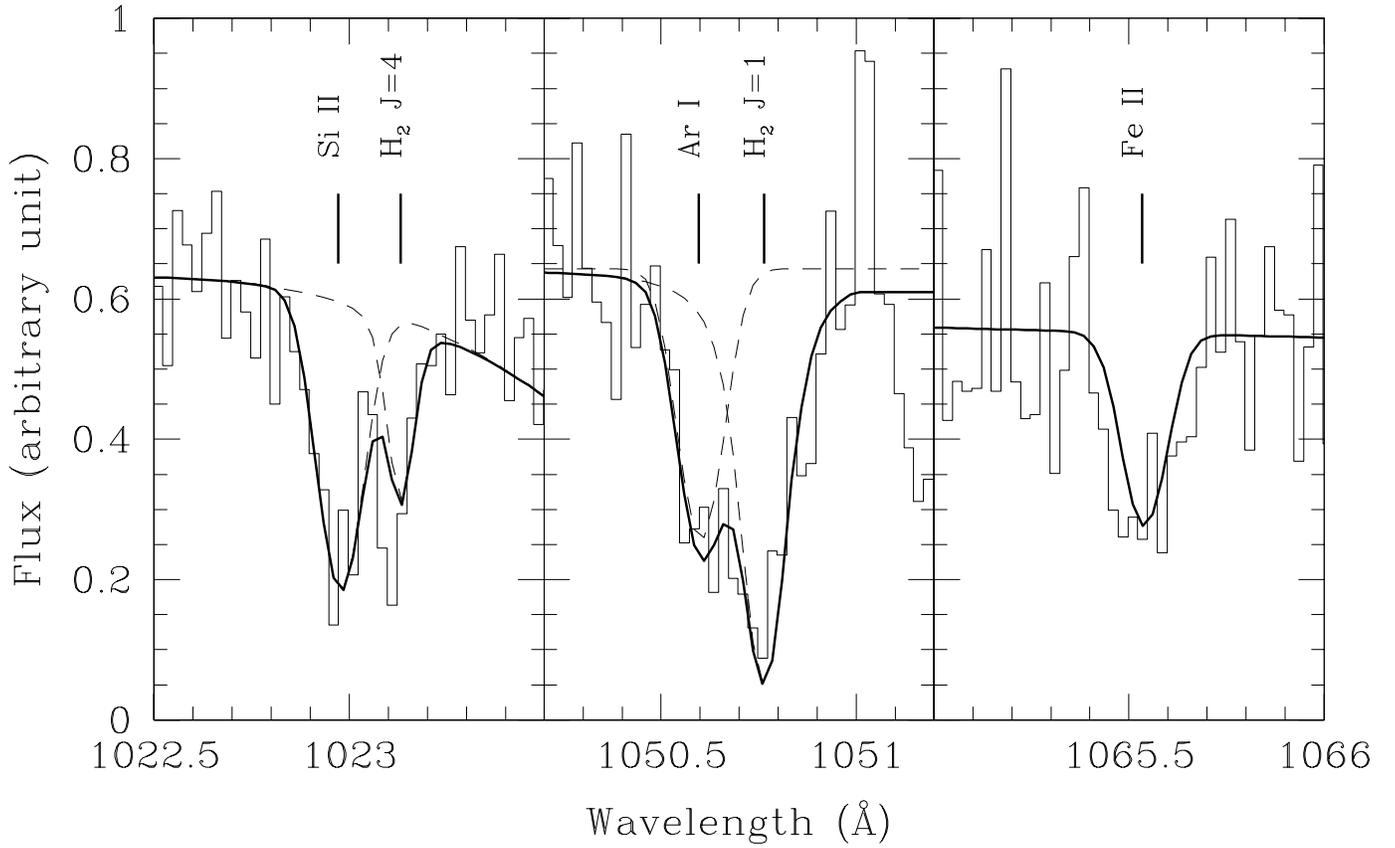}
\figcaption{Plot of some atomic lines detected at~$\sim$750\kms .
Absorption lines of Si\,{\sc ii}, Ar\,{\sc i} and Fe\,{\sc ii} 
are from \iz. 
The Si\,{\sc ii} and Ar\,{\sc i} lines are blended with Galactic H$_2$ lines.
This blend is easily resolved because the 
Galactic H$_2$ is detected in many other lines.
\label{atomic lines}}
\end{figure*}

The high velocity cloud at $-160$\kms\ is detected 
in C\,{\sc ii}, Cr\,{\sc ii}, Fe\,{\sc ii}, and Si\,{\sc ii} lines.
The second component identified with galactic clouds 
shows absorption lines not only from atoms and
ions (e.g., C\,{\sc ii}, O\,{\sc i}, Fe\,{\sc ii} and Si\,{\sc ii}) but 
also from molecular hydrogen.
Lines from H$_2$ at levels up to at least J=5 are detected. This is the
only component showing the presence of these electronic transitions of H$_2$.
This component shows a complex structure, suggesting the presence
of several interstellar clouds separated by up to 20~\kms ,
and will not be detailed further.
Finally the third component detected at 750~\kms\ is seen
in Ar\,{\sc i}, N\,{\sc i}, Fe\,{\sc ii}, and Si\,{\sc ii}
(Fig.~\ref{atomic lines}).

In addition to these three main components, 
a complex structure is observed around 1032~\AA\ and 1037~\AA\ .
This corresponds to the presence of O\,{\sc vi} lines with 
radial velocities between $-100$~\kms\ and $+150$~\kms\ originating in the
galactic halo.

\begin{figure*}
\plotone{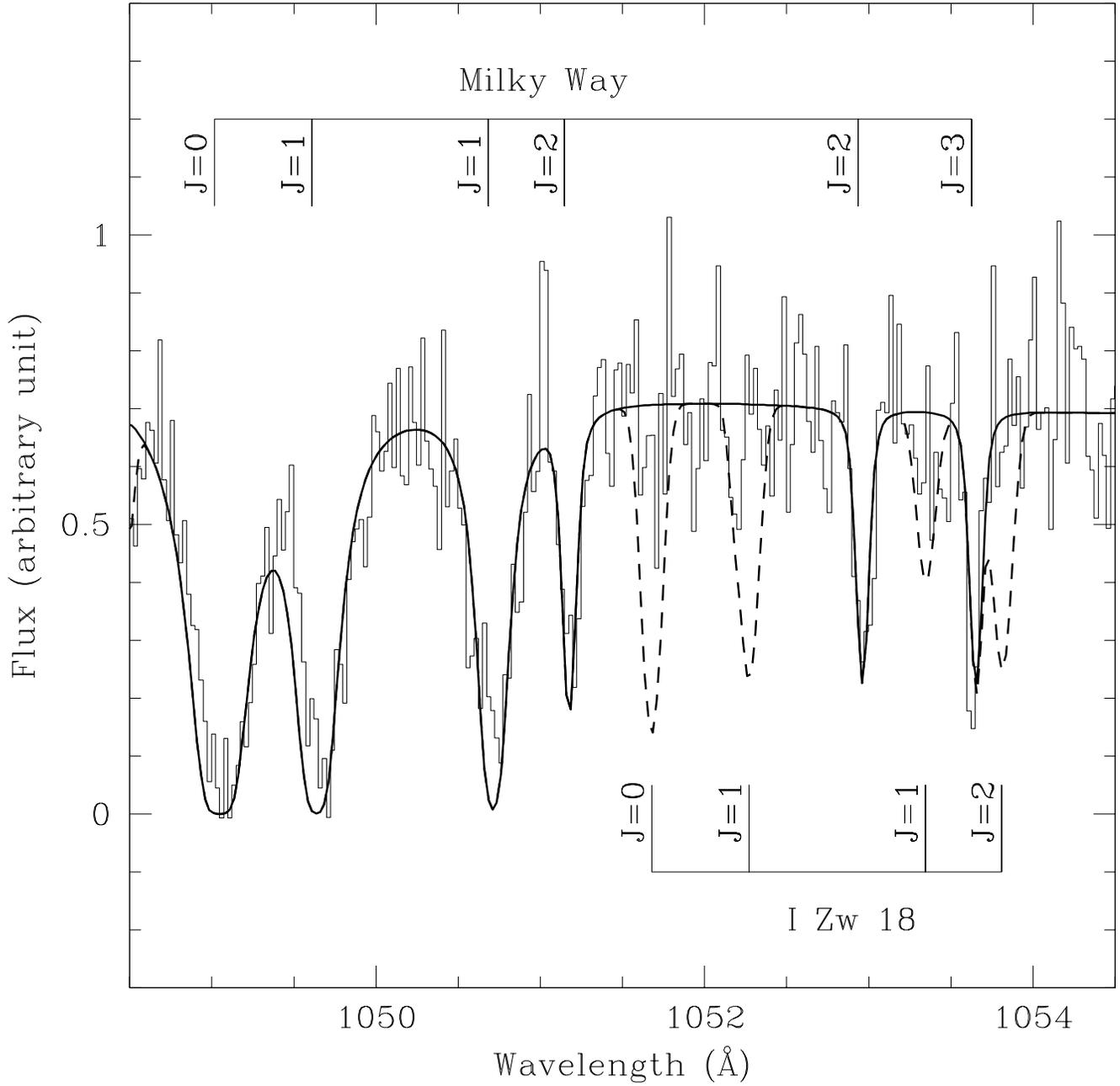}
\figcaption{Plot of the 4-0
H$_2$ Lyman bands.
Although the Galactic H\,{\sc i} column density ($\sim 10^{20}$~cm$^{-2}$) 
is lower than the one from \iz\ ($\sim 10^{21}$~cm$^{-2}$), 
the Galactic H$_2$ is easily detected ($\sim 10^{20}$~cm$^{-2}$). 
No line of the H$_2$ bands is detected at the radial velocity of \iz .
The dashed lines shows the expected lines if the column density
of H$_2$ had been $10^{15}$~cm$^{-2}$ in the plotted J levels.
\label{plot h2}}
\end{figure*}

No line from H$_2$ is observed
at the radial velocity of \iz\ (Fig.~\ref{plot h2}). 
We calculated the upper limits of
the H$_2$ column densities assuming an intrinsic width of the 
lines of $b=18$~\kms\ (van Zee et al. 1998). The limits have been estimated
by calculating the difference between a simulated spectrum and
the observed spectrum in 9 Lyman bands (0-0 to 8-0) and calculating
the corresponding increase of the $\chi^2$ of the fit to the spectrum.
The upper limits quoted in Table~\ref{H2 IZw18} give an increase of
the $\chi^2$ larger than 100, 
corresponding to a non detection at the $\sim 10 \sigma$~level.
These limits are thus very conservative and correspond to a total
column density $N_{\rm tot}({\rm H}_2)\la 10^{15}$~cm$^{-2}$.
A different intrinsic width of the lines would not change the result 
significantly.

The H\,{\sc i} Lyman $\beta$ line is strongly perturbed by the airglow lines.
However using only the blue wing of the absorption line
and assuming that this wing is due to the H\,{\sc i} of \iz\ at 750~\kms,
it is possible to obtain an estimate of the H\,{\sc i} column density.
We find N(H\,{\sc i})$\approx 2.1\times 10^{21}$cm$^{-2}$. This value 
is consistent with N(H\,{\sc i})$\approx 3.5\times 10^{21}$cm$^{-2}$
obtained with HST with a narrow slit (Kunth et al. 1994)
and the peak column density of 
$3.0\times 10^{21}$cm$^{-2}$ obtained with observations
of the 21~cm emission line.
Assuming a constant N(H$_2$)/N(H\,{\sc i}) ratio across the whole galaxy, 
we can scale the upper limit on H$_2$ column density to a limit for the 
total mass of diffuse H$_2$, yielding $M_{{\rm H}_2} \la 30 $M$_\sun$.

Other lines of atoms and ions are observed in the \iz\ system at 750~\kms.
For instance, lines of Ar\,{\sc i}, N\,{\sc i}, Fe\,{\sc ii}, and 
Si\,{\sc ii} are clearly detected.
Neither O\,{\sc vi}, nor the electronic transitions of CO are detected at
the \iz\ radial velocity. The lack of CO absorption lines is not surprising 
since, contrary to diffuse H$_2$, CO should be confined in very dense 
clouds opaque to UV sources. The problem of
O\,{\sc vi} will be discussed in a forthcoming paper.

\section{Discussion}

The interpretation of the lack of absorption lines of H$_2$ in the
spectrum of \iz\ deserves a 
detailed discussion. Note first that {\it FUSE} gives access to the average 
absorption over the full body of \iz , providing $\ga10^3$~lines of sight
to stars emitting in the far--UV and gathered in a central 
region approximately $10\arcsec$ wide.  Some of these stars are 
resolved by the HST (Dufour et al. 1996). Our observations are not sensitive
to dense molecular clouds since: i) dust, even in minute amounts will hide 
the background stars in the far--UV. ii) even if such clouds were 
transparent to UV photons, H$_2$ absorption lines would not be detected at 
our S/N ratio unless the covering fraction is larger than $\sim10\%$. On the 
other hand, our observations are very sensitive to diffuse H$_2$.
Its absence is very unusual.
Indeed in our Galaxy, H$_2$ is strongly detected for H\,{\sc i} 
column densities larger than a few $10^{20}$~cm$^{-2}$, and is often 
detected for lower N(H\,{\sc i}) (Dixon et al. 1998). 
With 2N(H$_2$)/N(H\,{\sc i})$\ll 10^{-6}$ and H\,{\sc i} column density as 
high as N(H\,{\sc i})$\approx 2\times 10^{21}$cm$^{-2}$, 
our observation is placed in the extreme bottom right corner of 
Fig.~5 of Dixon et al. (1998) 
representing the fraction of molecular hydrogen versus N(H\,{\sc i}).
Such an extreme situation has never been observed within a galaxy.
Even the Magellanic Clouds, with sub-solar metallicities and high far--UV 
radiation fields, show detectable H$_2$ along sightlines with lower 
H\,{\sc i} column densities (e.g., Friedman et al. 2000, Shull et al. 2000).
We also would like to stress that this result raises an 
interesting similarity with what is observed 
in the damped Ly$\alpha$ systems 
in QSO lines of sight at higher redshift. Despite their high levels of 
H\,{\sc i}, having low metalicity, low dust content and high UV
environment, they also present no detectable H$_2$ (Black et al. 1987).

We now show that the lack of H$_2$ in the diffuse ISM of \iz\
is a consequence of the low abundance
of grains, of the high ultraviolet flux and of the low atomic density in the 
H\,{\sc i} cloud surrounding \iz . 

There are two possible mechanisms for the formation of H$_2$ 
in the H\,{\sc i} cloud: formation via H$^-$ 
(see e.g. Jenkins \& Peimbert 1997), or
combination of two H atoms on a dust grain (Hollenbach \& Salpeter 1971).
A third mechanism involving the production of H$_2^+$ 
by radiative association
of H and H$^+$ is very inefficient in the present case since 
the reaction is slow 
and will not be considered further.
The first mechanism for H$_2$ formation starts with the formation of a
negative ion,
$
{\rm H} + e \rightarrow {\rm H}^- + h\nu,
$
with a rate $1.0\times 10^{-15}$ $T_3$ exp$(-T_3/7)$ cm$^3$ s$^{-1}$, $T_3$
being the temperature in units of 10$^3$ K (Jenkins \& Peimbert 1997). 
This is
followed by the faster associative detachment reaction H$^-$ + H
$\rightarrow$ H$_2$ + $e$. The electrons come mainly from the
photoionization of carbon: $n_e=$(C/H)\,$n$(H). 
We assume that the abundance of carbon in the H\,{\sc i}
cloud is the same as in the H\,{\sc II} region: 
C/H = $3.5\times 10^{-6}$ (Garnett et al. 1997).
The rate of formation is then as low as 
$\sim 10^{-20} n({\rm H})^2$~cm$^3$s$^{-1}$ at a
temperature of 10$^4$ K, the most favorable case, so that the mechanism is
very inefficient unless the medium contains clumps with very high densities.

The formation of H2 on grains is a more efficient mechanism if the dust
is cold enough for the H atoms to stick and remain on the
grain surface long enough to combine.
To estimate the grain temperature,
we examine what happens at the edge of the H\,{\sc i} cloud
of \iz , where the UV flux which photodissociates H$_2$ is minimal. 
The angular radius of the H\,{\sc i} cloud is approximately 
30\arcsec\ from the VLA map of van Zee et al. (1998), corresponding to a 
radius $R_0$ = 1.7 kpc
at the distance of \iz , taken as 11.5 Mpc from its radial velocity of
750 km s$^{-1}$ ($H_0 = 65$~km~s$^{-1}$~Mpc$^{-1}$).
The radiation flux from the ionizing stars of \iz\ around 1000~\AA\ 
measured by {\it FUSE} or extrapolated from IUE observations is
about $3\times 10^{-14}$~erg s$^{-1}$ cm$^{-2}$~\AA$^{-1}$.
Correcting for the Galactic extinction
($E(B-V) = 0.04$ mag., Kunth et al. 1994),
we obtain a UV flux at the Earth 
of approximately $4.5\times 10^{-14}$~erg s$^{-1}$ cm$^{-2}$~\AA$^{-1}$. 
This yields at $R_0$ a flux 
$F_{1000} \simeq 2\times 10^{-6}$ erg s$^{-1}$ cm$^{-2}$ \AA$^{-1}$.
We finally find the grain temperature at $R_0$ by solving the
temperature equilibrium equation
%
$
\pi a^2 \int Q_a(\lambda) F_{\lambda} \, d\lambda = 4 \pi a^2  \int
Q_a(\lambda) \pi B_{\lambda}(T) \, d\lambda,
$
%
where $a$ is the radius of the grain assumed to be spherical,
$Q_a(\lambda)$ its absorption efficiency, $F_{\lambda}$ the incoming UV flux
and $B_{\lambda}(T)$ the Planck function at the temperature T of the grain.
As any kind of grain is strongly absorbing in the far-UV which dominates
strongly the radiation field, we take $Q_a \simeq 1$ in the left side of
the equation. In the far-IR where the grains emit, we take 
$
Q_a(\lambda) \simeq  0.1 (\lambda/100\,\mu{\rm m}) ^{-2}
(a/0.1\,\mu{\rm m})
$
(Draine \& Lee 1984).
It is then possible to solve the temperature equation analytically, finding
at $R_0 : $
$
T \simeq 15.5 (a/0.1\,\mu{\rm m})^{-1/6} \; {\rm K}.
$
This grain temperature is close to that in the diffuse Galactic
interstellar medium and allows the formation of H$_2$. However $T$
increases as $R^{-1/3}$ closer to \iz\ and H$_2$ cannot form on grains in
the inner parts of the H\,{\sc i} cloud.

In the steady state, the molecular hydrogen is also destroyed through 
absorption in the Lyman bands, and the resulting H$_2$ column density
can be estimated.
We can take for the formation rate on grains ${\cal R}$ 
the canonical Galactic value of $10^{-17} n({\rm H})^2$~cm$^3$~s$^{-1}$  
(Hollenbach \& Salpeter 1971) divided by 50 since the dust-to-gas ratio is 
less than 1/50 of the Galactic value (Kunth et al. 1994).
In the present case where the H$_2$ electronic 
bands are optically thin the fraction of molecular hydrogen 
$f({\rm H}_2) = 2n({\rm H}_2)/[2n({\rm H}_2)+n({\rm H})]$ is (Jura 1974)

\begin{equation}
f({\rm H}_2) = 2 {\cal R} n({\rm H})/I,
\end{equation}

where $I$ is a photodissociation rate. 
Jura (1974) has calculated $I$ for different cases and we simply use his
estimate close to the O9.5V star $\zeta$ Oph, scaled  
to $F_{1000}$, the flux at 1000~\AA\ at the radius $R_0$ of \iz .
We will assume that the
cloud is spherical and uniform, in which case its density is $n({\rm H}) =
N({\rm HI})/R_0$ = 0.4 atom cm$^{-3}$, $N({\rm HI})$ 
being the column density we measure in front of \iz .
We obtain $f({\rm H}_2) \simeq 2\times 10^{-9}$. 
The abundance of H$_2$ is still smaller closer to \iz\ since
the UV flux is accordingly larger. Thus the calculated column density of
H$_2$ is

\begin{equation}
N({\rm H}_2) \la 1 \times 10^{-9}\,N({\rm HI}) 
 \approx 2\times 10^{12}\;{\rm mol.\,cm}^{-2},
\end{equation}

less than the observed upper limit by more than two orders of magnitude. 

We thus conclude that our observation shows that the diffuse ISM
surrounding \iz\ cannot be very inhomogeneous at large scales,
otherwise H$_2$ would have been observed.
However it cannot be excluded that this
medium contains molecular clouds and in particular
the kind of very dense, discrete molecular clumps proposed
by Pfenniger et al. (1994) to account for the dark matter in our Galaxy.
These clumps would escape detection since the associated absorption 
would be observed only
in front of stars which, although very numerous, have a very small total
surface coverage. However the suggestion of Lequeux \& Viallefond (1980)
that the dark matter seen dynamically in \iz\ is made of widespread diffuse
molecular hydrogen is no longer tenable after the present observations.

\acknowledgments
{\bf Acknowledgments.} 

This work is based on data obtained for the Guaranteed Time Team by the
NASA-CNES-CSA {\it FUSE} mission operated by the Johns Hopkins University.
Financial support to U.S. participants has been provided by NASA contract
NAS5-32985. We thank E. Roueff for providing H$_2$ transition data in 
electronic format.


\clearpage

\begin{deluxetable}{lll}
\tablecaption{Upper limits on the H$_2$ 
content of \iz\ at $\sim$10~$\sigma$ level
\label{H2 IZw18}}
\tablewidth{0pt}
\tablehead{
\colhead{Molecule} & \colhead{ $J$ }   & \colhead{$N$ }\\
\colhead{} & \colhead{}   & \colhead{ (cm$^{-2}$) }}
\startdata
H$_{2}$ & 0 & $<5\times 10^{14}$ \\
H$_{2}$ & 1 & $<6\times 10^{14}$ \\
H$_{2}$ & 2 & $<5\times 10^{14}$ \\
H$_{2}$ & 3 & $<5\times 10^{14}$ \\
H$_{2}$ & 4 & $<9\times 10^{14}$ \\
\enddata
\end{deluxetable}


\clearpage

\begin{figure*}
\plotone{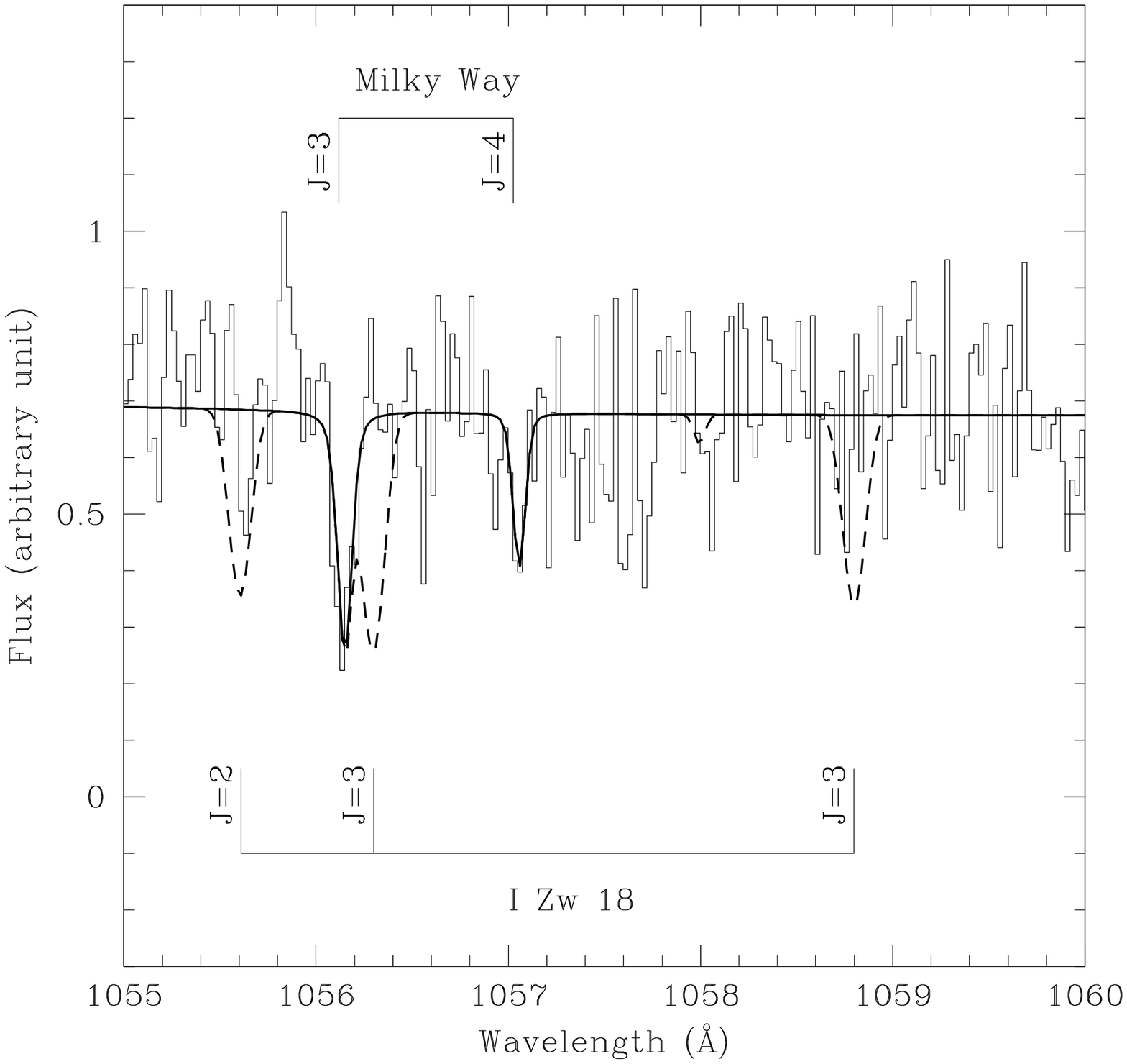}
\end{figure*}

\end{document}